\documentclass[twocolumn,showpacs,preprintnumbers,amsmath,amssymb]{revtex4}
\usepackage{graphicx}

\begin{document}

\title{Diffraction of a superfluid Fermi gas by an atomic grating}
\author{C. P. Search, H. Pu, W. Zhang, and P. Meystre}
\affiliation{Optical Sciences Center, The University of Arizona,
Tucson, AZ 85721}
\date{\today}

\begin{abstract}
An atomic grating generated by a pulsed standing wave laser field
is proposed to manipulate the superfluid state in a quantum
degenerate gas of fermionic atoms. We show that in the presence of
atomic Cooper pairs, the density oscillations of the gas caused by
the atomic grating exhibit a much longer coherence time than that
in the normal Fermi gas. Our result indicates that the technique
of a pulsed atomic grating can be a potential candidate to detect
the atomic superfluid state in a quantum degenerate Fermi gas.
\end{abstract}
\maketitle

Since the observation of Bose-Einstein condensation in trapped
atomic gases \cite{BEC} there has been increasing interest in the
possibility of observing the Bardeen-Cooper-Schrieffer (BCS) phase
transition to the superfluid state \cite{bardeen} in dilute
fermionic akali gases \cite{Li6,bohn,holland,mackie}. Currently,
experimental efforts in cooling of fermionic atoms of $^{6}$Li
\cite {truscott}and $^{40}$K \cite{jin} to the quantum degenerate
regime have made
significant progress, reaching temperatures as low as $0.2T_{F}$ where $%
T_{F} $ is the Fermi temperature.

At the ultracold temperatures achieved in these experiments,
$p$-wave collisions between atoms are highly suppressed and
$s$-wave collisions between atoms in the same internal state are
forbidden by the Pauli exclusion principle. As a result, the most
likely possibility for the formation of Cooper pairs is an
attractive $s$-wave interaction between atoms in different
hyperfine states. Fortunately, $^{6}$Li and $^{40}$K appear to be
very promising candidates. $^{6}$Li possesses an anomolously large
and negative $s$-wave scattering length \cite{Li6} while for
$^{40}$K, a Feshbach resonance exists for two of the hyperfine
states which can be used to create the required large attractive
interaction \cite{bohn}. There have been several recent proposals
involving the molecular state formed by either a Feshbach
resonance \cite{holland} or photoassociation \cite{mackie} which
would indicate that transition temperatures $\gtrsim 0.1T_{F}$
might be possible, thereby placing the BCS transition within reach
of current experiments.

Unlike Bose-Einstein condensation, the BCS transition is not
characterized by a significant change in either the density or
momentum distribution of the gas. Hence, how to identify the BCS
transition and to study the properties of the superfluid state in
these gases remain an open question. Recently, there have been a
number of proposals to detect the BCS state including off-resonant
light scattering \cite{zhang}, measuring the frequency of low
energy excitations \cite{baranov}, and the line shift in the
absorption spectrum of a resonant laser \cite{zoller}. Since the
Cooper pairing of the fermions is represented by a phase coherent
superposition of pairs of single particle states in the BCS wave
function, processes which involve the scattering of particles can
exhibit quantum interferences that are not present in the normal
state \cite{tinkham}. However, the previous proposals
\cite{zhang,baranov,zoller} are only sensitive to the existence of
a superfluid order parameter and/or the spectrum of quasiparticle
excitations of the gas and do not directly probe the quantum
coherence of the BCS state. In this letter, we study the
possibility of using atom optics techniques to directly study the
coherence presence in a gas of fermionic atoms that have undergone
the BCS transition.

In our proposal, a pulsed atomic grating is generated by a far
detuned standing wave optical field of frequency $\omega_{L}$ and
applied to the Fermi gas, similarly to the recent experiment with
a Bose-Einstein condensate by Deng {\it et al.} \cite {deng}. The
grating results in the scattering of atoms from states of momentum
$\hbar {\bf k}$ to $\hbar \left( {\bf k}+l{\bf q}\right) $ where
$\pm {\bf q}/2$ are the wave vectors of the travelling waves
forming the grating and $l$ is an integer. For $|{\bf q}|\ll 2
k_{F}$, there is a rapid decay of the density oscillations in the
normal state due to a dephasing caused by the range of energies
near the Fermi surface that are excited by the grating. In
contrast, for the BCS state, even though this dephasing also
initially reduces the amplitude of the density oscillations, small
amplitude oscillations nevertheless persist indefinitely. They
result from the constructive quantum interference for the creation
of pairs of quasiparticles by the grating and the large density of
states for low energy quasiparticles. The detection of such
persistent density oscillations would provide a clear indication
of the BCS state.

We consider a box of volume $V$ containing $N$ fermionic atoms
with density $n=N/V$. The gas is assumed to be at zero temperature
($T=0$). The atoms possess two degenerate ground states, $\left|
\uparrow \right\rangle $ and $\left| \downarrow \right\rangle $,
that have equal population and equal chemical potentials, $\mu $.
For convenience, we refer to the two states as spin states. The
fermion annihilation (creation) operators $\hat{c}_{{\bf k}\sigma
} (\hat{c}_{{\bf k}\sigma }^{\dagger })$
destroy (create) an atom in a state with momentum $\hbar {\bf k}$ and spin $%
\left| \sigma \right\rangle $ and obey the anti-commutation relations $%
\left\{ \hat{c}_{{\bf k}\sigma },\hat{c}_{{\bf k}^{\prime }\sigma ^{\prime
}}^{\dagger }\right\} =\delta _{{\bf k,k}^{\prime }}\delta _{\sigma ,\sigma
^{\prime }}$ and $\left\{ \hat{c}_{{\bf k}\sigma },\hat{c}_{{\bf k}^{\prime
}\sigma ^{\prime }}\right\} =0.$ The Hamiltonian for the system is given by $%
\hat{H}=\hat{H}_{0}+\hat{H}_{I}$ where $\hat{H}_{I}$ represents
the interaction of the atoms with the optical potential formed by the laser and $%
\hat{H}_{0}$ is the free Hamiltonian for the fermions. It includes
attractive $s$-wave interactions between atoms in opposite spin
states with scattering length $a<0$. By introducing the order
parameter for the BCS state,
\begin{equation}
\Delta =\frac{
4\pi \hbar^{2}|a|}{Vm}\sum_{{\bf k}}\left\langle \hat{c}_{-{\bf k}\downarrow }%
\hat{c}_{{\bf k}\uparrow }\right\rangle,
\end{equation}
$\hat{H}_{0}$ may be diagonalized by a canonical transformation
\cite{tinkham},
\begin{equation}
\hat{H}_{0}=E_{0}+\sum_{{\bf k}}\hbar \nu _{{\bf k}}\left( \hat{\alpha}_{%
{\bf k}\uparrow }^{\dagger }\hat{\alpha}_{{\bf k}\uparrow }+\hat{\alpha}_{-%
{\bf k}\downarrow }^{\dagger }\hat{\alpha}_{-{\bf k}\downarrow
}\right),  \label{BCS}
\end{equation}
where $E_{0}$ is the ground state energy and $\hat{\alpha}_{{\bf
k}\sigma }^{\dagger }$ ($\hat{\alpha}_{{\bf k}\sigma }$) are
quasiparticle creation (annihilation) operators given by,
\begin{subequations}
\label{ukvk}
\begin{eqnarray}
\hat{\alpha}_{{\bf k}\uparrow } &=&\cos (\theta _{{\bf k}}/2)\,\hat{c}_{%
{\bf k}\uparrow }-\sin (\theta _{{\bf k}}/2)\,\hat{c}_{-{\bf k}%
\downarrow }^{\dagger }\,, \\
\hat{\alpha}_{-{\bf k}\downarrow }^{\dagger } &=&\cos (\theta _{%
{\bf k}}/2)\,\hat{c}_{-{\bf k}\downarrow }^{\dagger }+\sin (\theta _{%
{\bf k}}/2)\,\hat{c}_{{\bf k}\uparrow \,.}
\end{eqnarray}
\end{subequations}
The energies of the quasiparticles are $\hbar \nu _{{\bf k}}=\sqrt{\xi _{%
{\bf k}}^{2}+\Delta ^{2}}$ while $\tan \theta _{{\bf k}}=\Delta /\xi _{{\bf k%
}}$. Here, $\xi _{{\bf k}}=\hbar ^{2}k^{2}/2m-\varepsilon _{F}$ is
the kinetic energy measured relative to the Fermi energy,
$\varepsilon _{F}=\hbar ^{2}k_{F}^{2}/2m=\mu +2\pi
\hbar^{2}|a|n/m$. The ground state of $\hat{H}_{0}$, $\left|
\Phi \right\rangle ,$ is the vacuum state for the quasiparticles, $\hat{%
\alpha}_{{\bf k}\sigma }\left| \Phi \right\rangle =0$. For $\Delta
=0$ one recovers the normal state of a Fermi gas subject to a
Hartree-Fock mean-field energy of $-2\pi \hbar^{2}|a|n/m$ for each
spin component.

At time $t=0$ a standing wave laser field is applied to the gas
for a duration $\tau $, coupling the atoms to an excited state
of frequency $\omega _{a}$ with a Rabi frequency $\Omega .$ For detunings $%
\delta =\omega _{L}-\omega _{a}$ large compared to the excited state linewidth and $%
\Omega $, the excited state may be adiabatically eliminated
resulting in an optical dipole potential $U_{0}\cos ^{2}({\bf
q\cdot x}/2)$ where $U_{0}=\hbar \Omega ^{2}/4\delta $. In the
Raman-Nath regime, the energy of the quasiparticles excited by the
potential may be neglected in comparison to the interaction
energy, i.e., $U_{0}\gg \hbar \nu _{{\bf k}_{F}+{\bf k}_{max}}$
where $\hbar {\bf k}_{max}=U_{0}\tau  {\bf q}$ is the maximum
momentum imparted by the potential in the time $\tau $
\cite{meystre}. For the condition $q\ll 2 k_{F}$ considered in
this paper, this requires $\tau \ll \sqrt{\left[ 1-\left( \Delta
/U_{0}\right) ^{2}\right] /\left( \hbar k_{F}q/m\right) ^{2}}$. In
this limit, the potential acts as an impulsive phase grating with
the second quantized form,
\begin{equation}
\hat{H}_{I}=-\hbar \lambda \delta (t)\left( \hat{\rho}_{{\bf q}}+\hat{\rho}%
_{-{\bf q}}\right) ,  \label{inter}
\end{equation}
where $\lambda =-\tau \Omega ^{2}/16\delta $ and
\begin{equation}
\hat{\rho}_{{\bf q}}=\sum_{ {\bf k},\sigma }\hat{c}_{{\bf
k+q}\sigma }^{\dagger }\hat{c}_{{\bf k}\sigma }
\end{equation}
is the ${\bf q}^{th}$ Fourier component of the atomic density.
Physically, Eq. (\ref{inter}) may be interpreted as the result of
an atom absorbing a photon from one of the travelling waves
followed by the stimulated emission of a photon into the other,
thereby resulting in a momentum transfer to the atom of $\pm {\bf
q}$.

We wish to examine how the atomic density,
\begin{equation}
\rho ({\bf x},t)=\frac{1}{V}\sum_{{\bf q}^{\prime }}e^{-i{\bf q}%
^{\prime }\cdot {\bf x}}\left\langle \Phi \right| \hat{\rho}_{{\bf q}%
^{\prime }}(t)\left| \Phi \right\rangle,  \label{dens}
\end{equation}
evolves in time following the application of the grating. We work
in the Heisenberg representation with the time dependence for the
operators given by
\begin{equation}
\hat{c}_{{\bf k}\sigma }(t)=e^{-i\lambda \left( \hat{\rho}_{{\bf q}}+\hat{%
\rho}_{-{\bf q}}\right) }e^{i\hat{H}_{o}t}\hat{c}_{{\bf k}\sigma }e^{-i\hat{H%
}_{o}t}e^{i\lambda \left( \hat{\rho}_{{\bf q}}+\hat{\rho}_{-{\bf q}}\right)
}.  \label{Heis}
\end{equation}
A direct application of Eq. (\ref{BCS}) to calculate $\hat{c}_{{\bf k}\sigma
}(t)$ for $t>0$ would, in general, lead to incorrect results because Eq. (%
\ref{BCS}) depends on the mean-field $\Delta $, which is modified
by the scattering off the optical potential and  must be
calculated in a self-consistent manner. For $t>0$, one must use
the new value of the order parameter to perform a canonical
transformation on $\hat{H}_{0}$. Let us define the pair
expectation value at the time $t=0^{+}$ immediately after the
application of the grating as $F_{{\bf k}^{\prime },{\bf
k}}^{(+)}=\left\langle \hat{c}_{-{\bf k}^{\prime }\downarrow
}(0^{+})\hat{c}_{{\bf k}\uparrow }(0^{+})\right\rangle$. Using the
result \cite{rojo},
\begin{equation}
e^{-i\lambda \left( \hat{\rho}_{{\bf q}}+\hat{\rho}_{-{\bf q}}\right) }\hat{c%
}_{{\bf k}\sigma }e^{i\lambda \left( \hat{\rho}_{{\bf q}}+\hat{\rho}_{-{\bf q%
}}\right) }=\sum_{s=-\infty }^{\infty }i^{s}J_{s}(2\lambda )\hat{c}_{{\bf k}%
-s{\bf q}\sigma }.  \label{bes}
\end{equation}
where $J_{s}$ is a Bessel function of integer order $s$, we have
the explicit form for $F_{{\bf k}^{\prime },{\bf k}}^{(+)}$,
\begin{equation}
F_{{\bf k}^{\prime },{\bf k}}^{(+)}=\frac{1}{2}\sum_{s,n=-\infty
}^{\infty }i^{n}J_{s}(2\lambda )J_{n-s}(2\lambda )\sin \theta _{{\bf k}-s{\bf q%
}}\delta _{{\bf k-k}^{\prime },n{\bf q}}.
\end{equation}
The mean-field following the grating, $\Delta ^{(+)}=\frac{4\pi \hbar^{2}|a|%
}{Vm}\sum_{{\bf k,k}^{\prime }}F_{{\bf k}^{\prime },{\bf k}}^{(+)}e^{-i({\bf %
k}^{\prime }-{\bf k})\cdot {\bf x}}$, is now spatially modulated at all
harmonics of ${\bf q}$. Since the probability that a pair state $({\bf k}%
\uparrow {\bf ,-k}\downarrow )$ is scattered by the grating into the pair
state $({\bf k}+n{\bf q}\uparrow {\bf ,-k}-m{\bf q}\downarrow )$ is $%
J_{n}^{2}(2\lambda )J_{m}^{2}(2\lambda )\sim \lambda ^{2(|n|+|m|)}$, for $%
\lambda <1$ the mean-field experienced by each atom will not
differ significantly from what it was for $t\leq 0.$ Therefore, we
restrict ourselves to $\lambda <1$ and use the zeroth order
approximation $\Delta ^{(+)}\approx \Delta $. In the examples
presented below, we have numerically calculated $F_{{\bf k},{\bf
k}+n{\bf q}}^{(+)}$
and confirmed that for $n \neq 0$, $F_{{\bf k},{\bf k}+n{\bf q}}^{(+)}/F_{%
{\bf k},{\bf k}}^{(+)}<0.1$.

Using Eq. (\ref{bes}), the scattering of the atoms at $t=0$ may be
treated to all orders in $\lambda $. After some algebra one
obtains an expression for the density,
\begin{widetext}
\begin{eqnarray}
\rho ({\bf x},t) &=&\rho _{o}+4\sum_{n=1}^{\infty }\sum_{s=-\infty
}^{\infty
}J_{s}(2\lambda )J_{n+s}(2\lambda )\left\{ f_{s,n}(t)\cos \left[ \left( {\bf q%
}\cdot {\bf x}+\pi /2\right) n\right] -g_{s,n}(t)\sin \left[ \left( {\bf q}%
\cdot {\bf x}+\pi /2\right) n\right] \right\} , \label{density}
\end{eqnarray}
\end{widetext}
\begin{widetext}
\begin{eqnarray}
f_{s,n}(t) &=&\frac{1}{V}\sum_{{\bf k}}\sin \frac{1}{2}\theta _{{\bf k}-s%
{\bf q}}\left\{ \cos \nu _{{\bf k}}t\cos \nu _{{\bf k}+n{\bf q}}t\sin \frac{1%
}{2}\theta _{{\bf k}-s{\bf q}}+\sin \nu _{{\bf k}}t\sin \nu _{{\bf k}+n{\bf q%
}}tA_{s,n}({\bf k)}\right\} ,\\
g_{s,n}(t) &=&\frac{1}{V}\sum_{{\bf k}}\sin \frac{1}{2}\theta _{{\bf k}-s%
{\bf q}}\left\{ \sin \nu _{{\bf k}}t\cos \nu _{{\bf k}+n{\bf q}}tB_{s,0}(%
{\bf k)}-\cos \nu _{{\bf k}}t\sin \nu _{{\bf k}+n{\bf q}}tB_{s,n}({\bf k)}%
\right\},
\end{eqnarray}
\end{widetext}
where $\rho _{o}$ is the spatially uniform background density, $A_{s,n}({\bf %
k)=}\sin \left[ \frac{1}{2}\theta _{{\bf k}-s{\bf q}}-(-1)^{s}\theta _{{\bf k%
}}-(-1)^{s+n}\theta _{{\bf k}+n{\bf q}}\right] +(-1)^{n}\sin \theta _{{\bf k}%
}\sin \theta _{{\bf k}+n{\bf q}}/\sin \frac{1}{2}\theta _{{\bf k}-s{\bf q}}$%
, and $B_{s,n}({\bf k)=}\sin \left[ \frac{1}{2}\theta _{{\bf k}-s{\bf q}%
}-(-1)^{s+n}\theta _{{\bf k}+n{\bf q}}\right] $. For $\Delta =0,$
we recover the results of Ref. \cite{rojo}.

\begin{figure}
\includegraphics*[width=6.7cm,height=7.2cm]{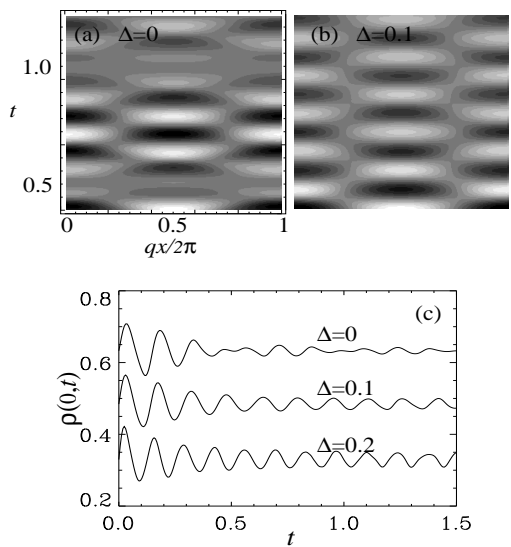}
\caption{Density oscillation of the Fermi gas after the grating.
(a), (b) The spatio-temporal evolution of the atomic density; (c)
density oscillations at $x=0$. The curves for $\Delta =0.1$ and
0.2 have been shifted down by 0.15 and 0.3, respectively. The
units for time is the so-called Talbot time\cite{rojo} $\tau_T =
2\pi/[\hbar q^2/(2m)]$, and the units for energy and density are
$\varepsilon_F$ and $k_F$, respectively. Here $\lambda=0.2$,
$q=0.3 k_F$. For (a) and (b), the brighter regions represent
higher density.} \label{fig1}
\end{figure}

Since the formation of Cooper pairs by the atoms only affects
those states lying in a shell of thickness $\delta k\approx
m\Delta /\hbar ^{2}k_{F}$ around $k_{F}$, ${\bf q}$ must be chosen
such that $q \sim \delta k$ in order to be sensitive to the
quantum coherence near the Fermi surface. This condition can be
reexpressed as $q/k_{F} \sim \Delta /\varepsilon _{F}$. For %
$^{6}$Li or $^{40}$K, realistic values of the zero temperature gap are $%
\Delta /\varepsilon _{F}\sim 0.04-0.1$ \cite{Li6,bohn}, although
recent models that account for the formation of a molecular
condensate formed using a Feshbach resonance or photoassociation
predict $\Delta /\varepsilon _{F}\sim 0.1-0.5$
\cite{holland,mackie}. Fig.~\ref{fig1} shows the density evolution
for various $\Delta$, with ${\bf q}$ along the $x$-axis and the
${\bf k}$-space of the fermions taken to be one-dimensional for
simplicity. We observe that the initial oscillations in the
density die out in a characteristic time $t_{d}$ for both the
normal state and superfluid state. This decay is attributable to
the single particle states near the Fermi surface, with the range
of energies $\delta E$ that are excited by the grating. The decay
time is therefore roughly given by $t_{d}\sim \hbar /\delta E$.
For $\lambda < 1$, ${\bf q}$ is the dominant spatial harmonic in
the density. Consequently, for the normal state, only those states
lying within a distance $q$ from the Fermi surface can be excited
by the grating due to Pauli blocking so that $ \delta E\approx
\left| \xi _{{\bf k}_{F}}-\xi _{{\bf k}_{F}-{\bf q} }\right| $ and
$t_{d}\sim m/\hbar qk_{F}.$ For the BCS state, those initial
states lying within the shell $\delta k$ of $k_{F}$ give the
greatest contribution when excited by the grating and therefore,
$\delta E \approx \hbar \left| \nu _{{\bf k}_{F}}-\nu _{{\bf
k}_{F}-{\bf \delta k}}\right| \approx 2\Delta $ and $t_{d}\sim
\hbar /2\Delta$. The ratio of dephasing times for the normal state
and the superfluid state is $\left( \Delta /\varepsilon
_{F}\right) (q/k_{F})^{-1}\sim 1$. This explains why the decay
times in Fig.~\ref{fig1}(c) appear to be the same.

The major difference between the superfluid state and the normal
state is the persistence of almost undamped density oscillations
for $t>t_{d}$ in the former case. From Fig.~\ref{fig1}, we see
that for the normal state, the density oscillations collapse at
times equal to $l\tau_T/2$ ($l$ is an integer) with incomplete
revivals between collapses. This is due to the discontinuity in
the momentum distribution of the normal state at $k_{F}$, which
leads to a dephasing term oscillating in time. For finite
temperatures, the momentum distribution is a smooth function of
${\bf k}$ in the normal state and hence revivals will not occur.
Such collapses and revivals are absent for the BCS state even at
$T=0$ since the momentum distribution in the ground state is a
smooth function of ${\bf k}$ for all ${\bf k}$. The origin of the
undamped oscillations may be understood by evaluating the density
to lowest order in perturbation theory,
\begin{eqnarray}
\rho ({\bf x},t)=\rho _{o}&+& \frac{4\lambda }{V}  \sum_{{\bf k}}
\sin
 ^{2}  \left[  \frac{1}{2}\left( \theta _{{\bf k}+{\bf q}}+\theta
_{{\bf k}}\right) \right] \nonumber \\
&& \sin \left[ \left( \nu _{{\bf k}+{\bf q}}+\nu _{{\bf k}}\right) t
\right] \cos ({\bf q}%
\cdot {\bf x}) .  \label{po}
\end{eqnarray}
The physical effect of the grating is to excite pairs of
quasiparticles with energies $\hbar \nu _{{\bf k}+{\bf q}}$ and
$\hbar \nu _{{\bf k}}.$ However, when Eq. (\ref{inter}) is
expressed in terms of the quasiparticle operators,
one finds that both $\hat{c}_{{\bf k+q}\uparrow }^{\dagger }\hat{c}_{{\bf k}%
\uparrow }$ and $\hat{c}_{-{\bf k}\downarrow }^{\dagger }\hat{c}_{-{\bf k}-%
{\bf q\downarrow }}$ contain the term $\hat{\alpha}_{{\bf k}+{\bf q}\uparrow
}^{\dagger }\hat{\alpha}_{{\bf k}\downarrow }^{\dagger }$ with the
respective amplitudes $\cos \left( \theta _{{\bf k}+{\bf q}}/2\right) \sin
\left( \theta _{{\bf k}}/2\right) $ and $\cos \left( \theta _{{\bf k}%
}/2\right) \sin \left( \theta _{{\bf k}+{\bf q}}/2\right) .$ As a result,
the transition amplitudes for the two processes in which an atom with spin $%
\left| \uparrow \right\rangle $ is scattered by the grating from
${\bf k}$ to ${\bf k+q}$ and an atom with spin $\left| \downarrow
\right\rangle $ is scattered by the grating from $-{\bf k}-{\bf
q}$ to $-{\bf k}$ add constructively to give a total probability
of creating a quasiparticle pair
proportional to $\sin ^{2}\left[ \left( \theta _{{\bf k}+{\bf q}}+\theta _{%
{\bf k}}\right) /2\right] .$ The constructive interference reaches
its maximum when
$\left| {\bf k}+{\bf q}\right| $ and $\left| {\bf k}\right| $ are within $%
\delta k$ of $k_{F}$ so that $\theta _{{\bf k}+{\bf q}}\sim \theta _{{\bf k}%
}\sim \pi /2.$ In contrast, for the normal state, the
quasiparticle pair corresponds to an atom excited above the Fermi
surface and a hole inside the Fermi sea and the terms
$\hat{c}_{{\bf k+q}\uparrow }^{\dagger }\hat{c}_{{\bf k}\uparrow
}$
and $\hat{c}_{-{\bf k}\downarrow }^{\dagger }\hat{c}_{-{\bf k}-{\bf %
q\downarrow }}$ are independent of each other. Alternately, one may note
that if $|{\bf k|}$ and $|{\bf k+q}|$ lie within $\delta k$ of $k_{F}$, then
the pair states $({\bf k}\uparrow {\bf ,-k}\downarrow )$ and $({\bf k}+{\bf q%
}\uparrow {\bf ,-k}-{\bf q}\downarrow )$ have formed Cooper pairs
in the superfluid state. Consequently the terms $\hat{c}_{{\bf
k+q}\uparrow }^{\dagger }\hat{c}_{{\bf k}\uparrow }$ and
$\hat{c}_{-{\bf k}\downarrow }^{\dagger }\hat{c}_{-{\bf k}-{\bf
q\downarrow }}$ are no longer independent and exhibit a quantum
interference due to the coherence between the single particle
states in each of the Cooper pairs.

\begin{figure}
\includegraphics*[width=6.cm,height=3.4cm]{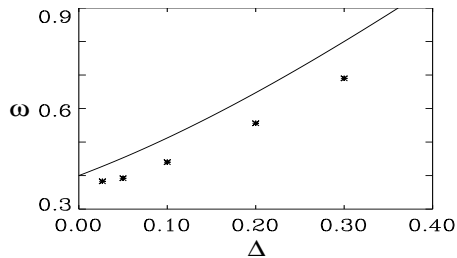}
\caption{Oscillation frequency of the density modulation as a
function of $\Delta$. the solid line is the analytical result (see
text) and the stars are numerical results obtained from
Eq.(\ref{density}). The units for frequency and energy are
$\varepsilon_F / \hbar$ and $\varepsilon_F$, respectively. Here
$\lambda=0.2$, $q=0.2 k_F$.} \label{fig2}
\end{figure}

In order to account for the degeneracy of the quasiparticle states
with energy $\varepsilon =\hbar \nu _{{\bf k}}$, the summation
over ${\bf k}$ may be replaced with an integration over the
quasiparticle energies. The density of states for quasiparticles
with energies in the interval $\left[ \varepsilon ,\varepsilon
+d\varepsilon \right) $ is $N_{s}(\varepsilon
)d\varepsilon =\varepsilon \left(\varepsilon ^{2}-\Delta ^{2}\right)^{-1/2}%
N_{n}(\xi )\Theta (\varepsilon -\Delta )d\varepsilon $ where
$N_{n}(\xi )$ is the density of states in the normal
state\cite{tinkham}. Since $N_{s}(\varepsilon )$ becomes very
large as $\varepsilon \rightarrow \Delta $ \cite {comment}, the
states which give the greatest contribution to Eq. (\ref{po}) are
those which lie on the Fermi surface. This corresponds to the
terms in Eq. (\ref{po}) with frequency $\nu _{{\bf k}_{F}+{\bf
q}}+\nu _{{\bf k}_{F}}$, which may be separated out from the sum.
The remaining states (with $\left| {\bf k}\right| \neq k_{F}$ ) in
the summation over ${\bf k}$, which get out of phase with each
other in a time on the order of $t_{d}$, lead to the collapse of
the initial large amplitude oscillations.

Consequently, the undamped oscillations are a result of the
combined effect of the constructive interference for the creation
of quasiparticles near the Fermi surface and the singular nature
of the density of states for quasiparticles with momentum near the
Fermi surface. A similar effect is observed in the relaxation rate
of polarized nuclear spins in superconductors, which shows a sharp
rise for temperatures just below $T_{c}$ \cite{tinkham}.

Figure~\ref{fig2} shows a comparison between $\hbar (\nu _{{\bf
k}_{F}+{\bf q}}+\nu _{{\bf k}_{F}} )\approx \Delta +\sqrt{\Delta
^{2}+4\varepsilon _{F}^{2}(q/k_{F})^{2}}$ for $q/k_{F}\ll 1$ and
the oscillation frequency obtained from the numerical evaluation
of Eq.~(\ref {density}) for $t\gg t_{d}$. Although there is some
discrepancy, the analytical result, $\nu _{{\bf k}_{F}+{\bf
q}}+\nu _{{\bf k}_{F}}$, does appear to predict the correct
dependence of the oscillation frequency on $\Delta$.

To put these results in perspective, we consider $^{40}$K with a
density of $n=2 \times 10^{14}$cm$^{-3}$ and $a=-53$nm
\cite{bohn}, which gives $k_{F}=(3 \pi^{2} n)^{1/3}=1.8\times
10^{7}$m$^{-1}$. The condition $|{\bf q}|\ll 2k_{F}$ then
corresponds to a laser wavelength satisfying $\lambda_{L}\gg
350$nm. Since the spatial extent of the trapped Fermi gases is
typically about $0.5$mm \cite{truscott}, the gas can be treated as
locally homogenous for a laser in the near infrared. Using
$\Delta=0.49\varepsilon_{F}\exp(-\pi/2k_{F}|a|)$ \cite{lifshitz},
we have $\Delta=0.095\varepsilon_{F}$. This gives a period for the
undamped oscillations, $T_{\Delta}\sim h/2\Delta$, of around
$0.1$ms which is much less than the typical life time of the
trapped gas.

In conclusion, we have shown that the manipulation of a degenerate
Fermi gas using atom optical techniques will be a sensitive probe
of the quantum coherence of the BCS state.

This work is supported in part by the US Office of Naval Research
under Contract No. 14-91-J1205, by the National Science Foundation
under Grant No. PHY98-01099, by the US Army Research Office, by
NASA, and by the Joint Services Optics Program.


\begin{references}
\bibitem{BEC}  M. H. Anderson {\it et al.}, Science {\bf 269}, 198 (1995);
K. B. Davis {\it et al.}, Phys. Rev. Lett. {\bf 75}, 3969 (1995);
C. C. Bradley {\it et al.}, Phys. Rev. Lett. {\bf 75}, 1687
(1995).

\bibitem{bardeen}  J. Bardeen {\it et al.}, Phys. Rev. {\bf 108}, 1175 (1957).

\bibitem{Li6}  H. T. C. Stoof {\it et al.}, Phys. Rev. Lett {\bf 76}, 10
(1996); M. Houbiers {\it et al.}, Phys. Rev. A {\bf 56}, 4864
(1997).

\bibitem{bohn}  J. L. Bohn, Phys. Rev. A {\bf 61}, 053409 (2000).

\bibitem{holland}  M. Holland {\it et al.}, Phys. Rev. Lett. {\bf 87},
120406 (2001).

\bibitem{mackie}  M. Mackie {\it et al.}, Opt. Express {\bf 8}, 118
(2000); M. Mackie {\em et al.}, e-print physics/0104043.

\bibitem{truscott}  A. G. Truscott {\it et al.}, Science {\bf 291}, 2570
(2001); F. Schreck {\it et al.}, Phys. Rev. Lett. {\bf 87}, 080403
(2001).

\bibitem{jin}  B. DeMarco and D. S. Jin, Science {\bf 285,} 1703 (1999); B.
DeMarco {\it et al.}, Phys. Rev. Lett. {\bf 86}, 5409 (2001).

\bibitem{zhang}  W. Zhang {\it et al.}, Phys. Rev. A {\bf %
60}, 504 (1999); F. Weig and W. Zwerger, Europhys. Lett. {\bf 49},
282 (2000).

\bibitem{baranov} M. A. Baronov and D. S. Petrov, Phys. Rev. A
{\bf 62}, R041601 (2000); G. Bruun and C. Clark, J. Phys. B {\bf
33}, 3953 (2000); M. A. Baranov, JETP Lett. {\bf 70}, 396 (1999).

\bibitem{zoller}  P. T\"{o}rma and P. Zoller, Phys. Rev. Lett. 85, 487
(2000).

\bibitem{tinkham}  M. Tinkham, {\it Introduction to Superconductivity}
(McGraw-Hill, New York, 1996).

\bibitem{deng}  L. Deng {\it et al.}, Phys. Rev. Lett. {\bf 83}, 5407 (1999).

\bibitem{meystre}  P. Meystre, {\it Atom Optics}(Springer, NY, 2001).

\bibitem{rojo}  A. G. Rojo {\it et al.}, Phys. Rev. A {\bf %
60}, 1482 (1999).

\bibitem{comment}  The inclusion of a quasiparticle lifetime will
remove the mathematical singularity in $N_s (\varepsilon)$ at the
gap.

\bibitem{lifshitz} E. M. Lifshitz and L. P. Pitaevskii,
{\it Statistical Physics Pt. 2 } (Pergamon, Oxford, 1980).
\end{references}
\end{document}